\documentclass[useAMS,usenatbib]{mn2e}
\usepackage{graphicx}
\title[Radius of baryonic collapse]{The radius of baryonic collapse in disc galaxy formation}

\author[S. Kassin et al.]{
Susan A. Kassin,$^1$\thanks{NASA Postdoctoral Program Fellow}\thanks{E-mail: susan.kassin@nasa.gov}  
Julien Devriendt,$^2$
S. Michael Fall,$^3$
Roelof S. de Jong,$^4$
\newauthor 
Brandon Allgood,$^{5,6}$ \&
Joel R. Primack$^{5}$ \\
$^1$ Astrophysics Science Division, Goddard Space Flight Center, Code 665, Greenbelt, MD 20771, USA\\
$^2$ Sub-Department of Astrophysics, University of Oxford, Denys Wilkinson Building, Keble Road, Oxford OX1 3RH, UK\\
$^3$ Space Telescope Science Institute, 3700 San Martin Drive, Baltimore, MD 21218, USA\\
$^4$ Astrophysikalisches Institut Potsdam (AIP), An der Sternwarte 16, 14482 Potsdam, Germany\\
$^5$ Department of Physics, University of California, Santa Cruz,1156 High Street, Santa Cruz, CA 95064, USA\\
$^6$ currently at: Numerate Inc., 1150 Bayhill Drive, San Bruno, CA 94066, USA
}

\begin{document}

\maketitle

\begin{abstract}
In the standard picture of disc galaxy formation, baryons and dark matter receive the same
tidal torques, and therefore approximately the same initial specific angular momentum.
However, observations indicate that disc galaxies typically have only about half as much 
specific angular momentum as their dark matter haloes.
We argue this does not necessarily imply that baryons 
lose this much specific angular momentum as they form galaxies.  It may instead indicate that galaxies are most directly related to the inner
regions of their host haloes, as may be expected in a scenario where baryons in the inner parts of haloes
collapse first.  A limiting case is examined under the idealised assumption of perfect angular momentum 
conservation.  Namely, we determine the density contrast $\Delta$, with respect to the critical density of the Universe,
by which dark matter haloes need to be defined in order to have the same average specific angular momentum as the galaxies they host.
Under the assumption that galaxies are related to haloes via their characteristic rotation velocities, the necessary $\Delta$
is $\sim 600$.  This $\Delta$ corresponds to an average halo radius and mass which are $\sim60$\% and
$\sim75$\%, respectively, of the virial values (i.e.,  for $\Delta = 200$).  We refer to this radius as the radius of baryonic collapse $R_{BC}$, 
since if specific angular momentum is conserved perfectly, baryons would come from within it.
It is not likely a simple step function due to the complex
gastrophysics involved, therefore we regard it as an effective radius.
In summary, the difference between the predicted initial and the observed final specific angular momentum of 
galaxies, which is conventionally attributed solely to angular momentum loss, can more naturally
be explained by a preference for collapse of baryons within $R_{BC}$, with possibly
some later angular momentum transfer.  
\end{abstract}


\section{Introduction}

In the standard picture of disc galaxy formation \citep[e.g.,][]{fall, dalc, mmw}, galaxies consist of a dissipative
baryonic component and a non-dissipative dark matter component.  Galaxies form hierarchially, and in this
process, baryons and dark matter acquire the same specific angular momentum ($j$) via tidal-torques.  This is 
because tidal-torques are most effective in the linear and the trans-linear regimes, when baryons and dark matter are well-mixed.  The 
dark matter then collapses non-dissipatively, and the baryons dissipatively, likely with some cloud-cloud
collisions and possibly shocks (processes which are expected to rearrange $j$ but not remove it).  
The baryons form rotating centrifugally-supported
discs at the centres of the potential wells.  For a review of this scenario see \citet{fall2}.  
This standard picture is  able to correctly predict galaxy properties such as scale-lengths and sizes
if the baryons retain most of their initial $j$.  It has been extended to 
include additional physics effects and larger samples of galaxies by e.g.,  \citet{fren},  \citet{cole},  
\citet{some}, \citet{dejo}, \citet{v2001}, \citet{hatt}, and \citet{dut9}.

In order for this scenario to correctly predict galaxy properties, the baryons must retain a large fraction of their initial angular momentum.  
However, early numerical simulations of galaxy formation contradicted this expectation \citep{katz, nav2, navwh}.  They found
a factor of $\sim30$ loss of angular momentum for simulated galaxies, and referred to this as an ``angular momentum catastrophe."  
As simulations improved over the years,  it became clear that much of this catastrophe  was 
actually a numerical artifact: too little resolution and too much numerical viscosity \citep[see e.g.,][and references therein]{gov10, alys, kere11, chris11, kimm}.  Another possible contribution to solving the angular momentum problem may be through feedback effects which
can delay baryons from falling onto discs \citep[e.g.,][]{weil, slars, eke, thac}.
With high numerical resolution and some feedback, galaxy simulations are now at a stage where angular momentum loss
may be a relatively minor problem.  In this paper, we explore another option: that the discs of galaxies draw baryons
mainly from the inner parts of dark matter haloes.  Some of the baryons in the outer parts may have not yet
collapsed onto the discs.

The angular momentum catastrophe prompted comparisons of the $j$ of simulated {\it haloes} 
to that of observed galaxies.  In these studies, the $j$ of 
dark matter haloes is measured out to the virial radius, $R_{Vir}$, which is standardly defined as $R_{\Delta=200}$, 
and is the effective radius at which the dark matter ceases to collapse into the halo.
\citet{stein} and \citet{burk} found that observed galaxies have 45\% and 70\% of the $j$ of their expected
host haloes in simulations, respectively, under the assumptions that galaxies can be related to 
simulated host haloes via characteristic rotation velocities directly and via a scaling factor, respectively.  
Recently, \citet{dutt11} found that the spin parameters of observed galaxies are $\sim 60$\% of those of simulated haloes.  These studies are consistent once differences in assumptions and approximations are accounted for.  

Studies which compare the total $j$ predicted for haloes by numerical simulations to that observed for galaxies all assume 
that the effective outer halo radius from which the baryons
collapse (defined here as $R_{BC}$) is equal to $R_{Vir}$.  Because baryons in the inner 
parts of haloes will have higher cooling rates and more frequent cloud-cloud collisions, it is reasonable to 
expect that they form the galaxies, and that baryons from larger radii are not captured.  Although $R_{Vir}$ has traditionally been identified with $R_{BC}$, 
these two radii are governed by different physics (dissipative versus non-dissipative), and need not be related, as emphasized by \citet{fall2}.
The only requirement is that $R_{BC}$ must be interior to $R_{Vir}$, since baryons cannot collapse
from unvirialized regions.  The purpose of this paper is to determine the effect of relaxing the 
assumption that $R_{Vir}$ and $R_{BC}$ are equal on the difference in $j$
between galaxies and haloes.  We assume for simplicity that the boundary between the 
collapsed and uncollapsed baryons is a sharp one.  In reality, it will be a gradual boundary because
some of the baryons in the halo within $R_{BC}$ might not collapse, and some baryons outside of $R_{BC}$ might.
Therefore, we regard  $R_{BC}$ as the effective boundary between these two regions.

In this paper, we ask the following question: If galaxies formed from all the baryons in haloes out to $R_{BC}$,
and beyond this radius the baryons remained in the halo, what is the value of $R_{BC}$ required to
match the $j$ of galaxies?  
We address this question by comparing the $j$ observed for disc galaxies with that of their expected dark matter haloes 
measured within a range of halo radii.  For disc galaxies, $j$ can be measured from observations of surface brightness 
profiles and rotation curves.  For dark matter haloes, we must resort to numerical simulations. 


This paper is organised as follows:  In \S2, we measure $j$ of dark matter haloes in a 
cosmological dark matter-only simulation.  We investigate its dependence on the halo radius within which $j$ is measured
and the halo radius at which the rotation velocity is measured.  The resulting predictions of dark matter halo $j$ are compared
to $j$ measured for a large observational sample of local galaxies for which the completeness is known in \S3.
A discussion of the results is in \S4.  We adopt a $\Lambda$CDM concordance universe ($\Omega_m=0.24, \Omega_{\Lambda}=0.76,
h=H_0/[100$ km s$^{-1}$ Mpc$^{-1}] = 0.73, \sigma_8 = 0.77, n = 0.958),$ i.e.,
within one standard deviation of both the WMAP 3 and 5 year best estimates \citep{sper, dunk}.
All logarithms are base ten.

\section{N-body simulation of dark matter haloes}

\begin{figure*}
\includegraphics[scale=1.05]{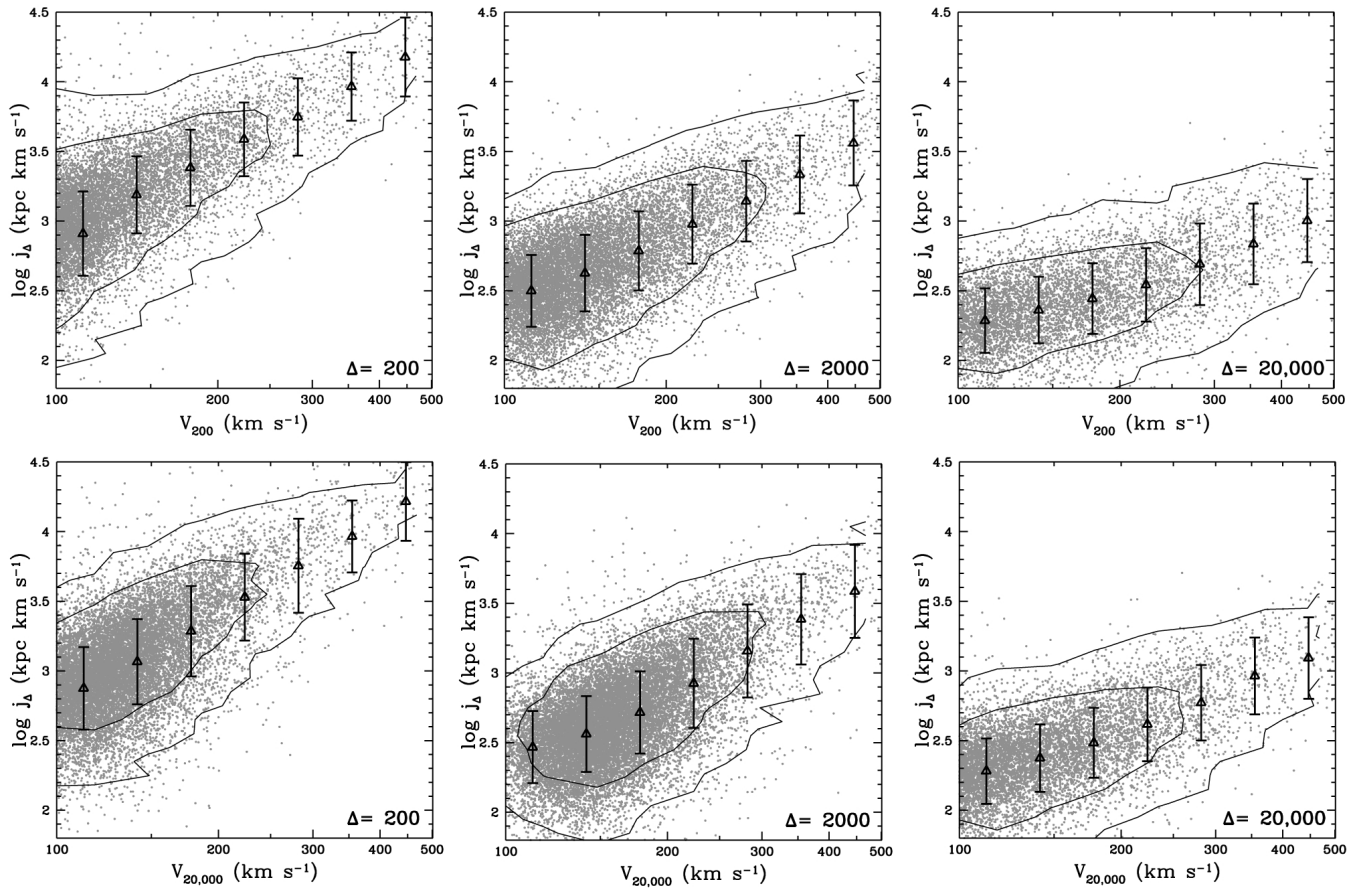}
\caption{For simulated dark matter haloes at $z=0$, the relations between $j_{\Delta}$ (for $\Delta = $ 200, 2000, and 20,000)
and rotation velocities $V_{200}$ and $V_{20,000}$ are shown.
Individual haloes are plotted as grey points, binned averages are shown as black triangles,
and the rms scatter is shown as black error bars.  Contours in volume density are shown for 2 and $20 \times 10^{-5}$
haloes per 0.1 in log $j_{\Delta}$ and per 0.1 log $V$, per Mpc$^3$.  The shapes of the distributions
are similar for $j_{\Delta}$ whether $V_{200}$ or $V_{20,000}$ is adopted.
As $\Delta$ increases, the normalisation of the relation between $j_{\Delta}$ and $V$ 
decreases, but the slope and scatter do not change greatly.  Similar relations are found for $V_{2000}$, 
but are not shown to avoid redundancy.
\label{fig:haloes_only}}
\end{figure*}


To quantify the dependence of dark matter halo $j$ on how the outer radius of a halo is
defined, we look to a suite of cosmological N-body simulations of dark matter haloes.  These
simulations include only dark matter and gravity (i.e., neither baryons nor hydrodynamics).  
As discussed in the Introduction, if the baryons in a given dark matter halo are
initially distributed in the same manner as the dark matter, and they later cool to form a  disc while conserving $j$,
then the $j$ of the galaxy should be equal to that of the virialized region of the dark matter halo.  However,
if baryons collapse progressively from the inner to the outer parts of haloes, and they have not finished
collapsing (or, if some baryons never collapse), then galaxy $j$ may be expected to reflect that of dark matter haloes within a given radius, $R_{BC}$.

To predict the distribution of $j$ among dark matter haloes, a large N-body simulation is needed which can model
the acquisition of angular momentum for even the slowest rotating galaxies
in our sample (125 km s$^{-1}$; \S3).  The simulation we adopt is part of the Horizon Project suite (http://www.projet-horizon.fr).  
This follows the evolution of a cubic cosmological volume of 100 $h^{-1}$ Mpc on a side (comoving)
containing $\sim 134$ million dark matter particles ($512^3$).  It starts at $z=99$ and is evolved using the publicly
available treecode Gadget 2 \citep{spri} with a softening length of 5 h$^{-1}$ kpc (co-moving).  The adopted cosmology results in a dark matter 
particle mass of $6.83 \times 10^8 M_{\odot}$.  Dark
matter haloes and the subhaloes they contain are identified with the {\tt AdaptaHOP} algorithm \citep{aube}.
The halo centres are positioned on the densest dark matter particle
located in the most massive substructure (see \citealt{twee} for details).
The total number of haloes and subhaloes in the simulation volume at $z=0$ 
with more than $100$ particles within $R_{200}$ and with circular velocities at this radius
which are greater than $100$ km s$^{-1}$ is 9661.  

The $j$ of a halo is measured within a range of radii as follows.  First, the halo is divided into 100 radial ellipsoidal shells, where
the axis ratios of the ellipsoid are obtained by computing the inertial tensor of all the particles in the halo.
Halo circular radii are defined as the cube root of the radii of the three major axes of each ellipsoid.
Next, the vector angular momentum of the particles in each shell is calculated, and the angular 
momenta of the shells is summed vectorially from the inner-most shell to the radii specified before taking its modulus.  The mass of a halo is
measured in an analogous manner, and $j$ is simply the angular momentum divided by the 
mass within a given radius.  Selected radii, $R_{\Delta}$, are defined by the density of the haloes 
with respect to the critical density of the universe ($\Delta \equiv \bar{\rho}(r<R_{\Delta}) / \rho_{\rm crit}$).
Specific angular momenta measured within these radii are defined as $j_{\Delta}$.
Circular velocities at these radii are $V_{\Delta} = (GM_{\Delta}/R_{\Delta})^{1/2}$, where 
$M_{\Delta}$ and $R_{\Delta}$ are the mass and radius of the halo defined by $\Delta$, and $G$ is the gravitational constant.
The ranges of $\Delta$, $R_{\Delta}/R_{200}$, and $M_{\Delta}/M_{200}$ probed are 50--20,000, 1.70--0.09, and 1.24--0.13, respectively.

\begin{figure*}
\includegraphics[scale=1.05]{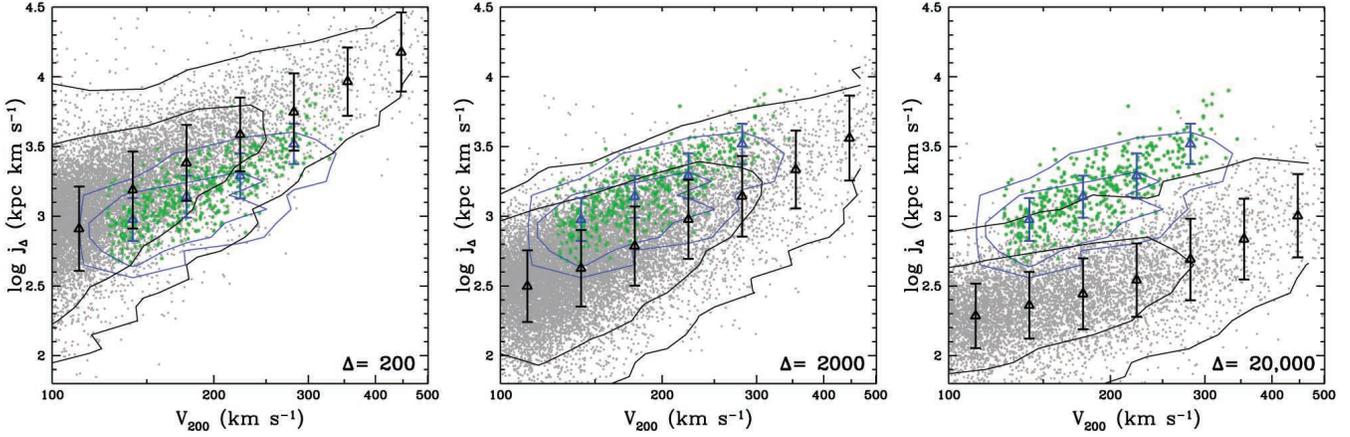}
\caption{These plots are the same as in Figure~\ref{fig:haloes_only} for $V_{200}$, except here observed galaxies 
are also shown ($V_{flat}$ of the galaxies is adopted as a characteristic rotation velocity and is plotted on the
horizontal axis).  
Individual galaxies are plotted as green points, and contours in volume density are shown in blue for 2 and $20 \times 10^{-5}$
galaxies per 0.1 in log $j_{\Delta}$ and per 0.1 in log $V_{200}$ per Mpc$^3$.  Binned averages for the galaxies
are shown as blue triangles, and the rms scatter is denoted by error bars.  
The scatter in $j$ for galaxies is about half of that of the haloes.
Under the assumption that characteristic galaxy and halo rotation velocities are equal (i.e., $V_{flat} = V_{200}$),
galaxies have on average a factor of $\sim 2$ less $j$ than haloes defined with $\Delta = 200$, a factor of $\sim2$ more
$j$ than haloes defined with $\Delta = 2000$, and a factor of $\sim 5.6$ more $j$ than haloes defined with $\Delta = 20,000$.  
\label{fig:haloes_gals}}
\end{figure*}

In Figure~\ref{fig:haloes_only}, relations between halo $j_{\Delta}$ and $V_{\Delta}$ are shown.\footnote{There 
is a drawback to a plot of $j$ versus $V$, namely
both axes incorporate factors of $V$, and a relation is expected by construction \citep[e.g.,][]{free}.
Because the local relation between galaxy $V$ and stellar mass is
tight \citep[e.g.,][]{bdej, kas2}, there is a similarly tight relation between between $j$ and stellar mass \citep[e.g.,][]{fall83},
which is not expected by construction.} 
Halo $j$ is measured within $R_{200}, R_{2000}$, and $R_{20,000}$, and halo $V$ is measured 
at $R_{200}$ and $R_{20,000}$.  We do not show results for halo $V$ measured at $R_{2000}$
since they do not differ significantly from those for $R_{200}$ or $R_{20,000}$.  
The radii $R_{2000}$ and $R_{20,000}$ correspond to 34\% 
and 9\% of $R_{200}$, respectively, on average  Only haloes with more than 100 particles are retained,
except for measurements of $j_{20,000}$ and $V_{20,000}$ for which haloes with more than 
50 particles are used.  For these 50-particle haloes, the intrinsic relations remain the same,
but the scatter is increased slightly due to increased Poisson noise.
The shapes of all the distributions are similar in terms
of slope and scatter, and are therefore approximately independent of the radius for which $j$ or $V$ are measured.
The slope flattens slightly with increasing $\Delta$, and the scatter remains about the same.
We will quantify this in the following section.
However, the normalisation is strongly dependent on $\Delta$: it decreases
by factors of $\sim 3$ and $\sim 6$  for 10 and 100-fold increases in $\Delta$, respectively.
A decreasing normalisation with increasing $\Delta$ is a consequence of how angular momentum is distributed in galactic haloes, 
with most of the angular momentum located in the outer parts.  As we increase $\Delta$, we exclude more and more
of the outer parts of the haloes, and the angular momenta decrease, as illustrated by
the simple analytic treatment in \citet[][Section 4]{fall83}.  In this paper, we quantify this 
decrease more precisely using numerical simulations.

\section{Comparison with observations of disc galaxies}

\begin{figure*}
\includegraphics[scale=1.0]{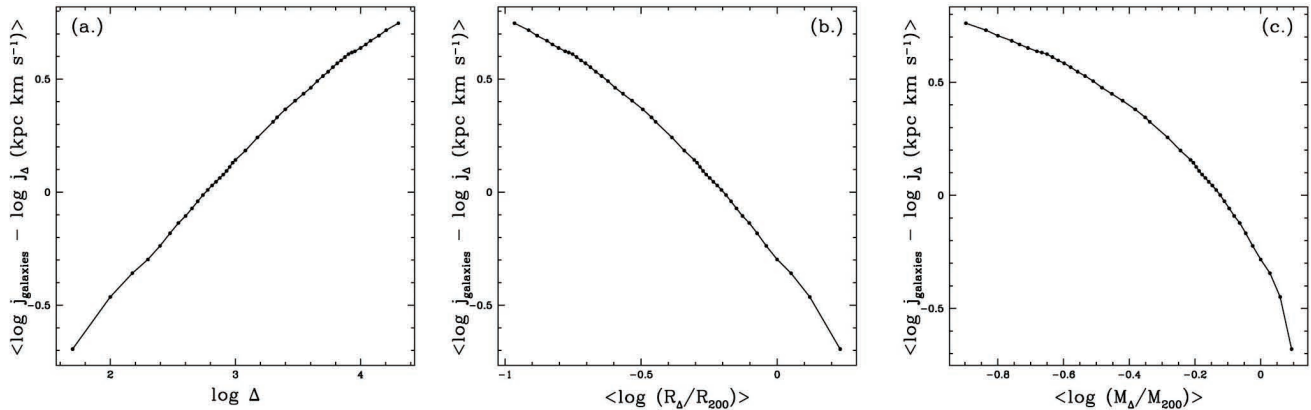}
\caption{The average difference between galaxy and halo log $j$, $ < \rm log \ j_{\rm galaxies}$ - log $j_{\Delta} >$,
is shown as a function of $\Delta$, $R_{\Delta}/R_{200}$, and $M_{\Delta}/M_{200}$,
in panels a, b, and c, respectively.  
Points demarcate discreet values, and solid lines simply connect the points.
There is no offset between galaxy and halo log $j$ for $\Delta = 578^{+34}_{-31}$, which corresponds to $R_{BC}=R_{\Delta=578}/R_{200} = 0.63^{+0.02}_{-0.01}$,
and $M_{\Delta=578}/M_{200} = 0.74 \pm 0.1$.
\label{fig:offset}}
\end{figure*}

The goal of this section is to place measurements for disc galaxies on Figure~\ref{fig:haloes_only}.  To do so,
we need to (1) adopt a galaxy sample for which the completeness is well-defined and which has
the necessary data available to derive circular velocities and
$j$, and (2) relate galaxies to simulated host dark matter haloes.  

To address the first need, a large sample of 456 galaxies from \citet{mat92} and completeness 
measurements from \citet{dejo} are adopted.  Details of this sample are given below.
The large size of and the data available for the sample necessitates simple estimates of 
$j$.  Therefore, we estimate $j$ as $2 V_{\rm flat} r_d$, where $V_{\rm flat}$ is the rotation velocity on 
the flat part of the rotation curve and $r_d$ is the scale-length of the galaxy disc.  This approximation
is exact for an exponential disc and a flat rotation curve.  Uncertainties in estimates of 
$j$ are $\sim 15$\%, which are dominated by errors in measurements of $r_s$ (mainly due 
to errors in sky background subtraction) and galaxy distances.
There are two minor effects on estimates of $j$, which we do not take into account,
but which work in opposite directions.
On the one hand, galaxies have rising rotation curves in their centres, and this causes the formula to slightly overestimate $j$.
On the other hand, most galaxies are expected to have extended gas discs, but with very little mass, which
would cause the formula to slightly underestimate $j$. 

The galaxy sample used is a sub-sample of the ESO-Uppsala Catalog of Galaxies \citep{laub} which was selected by eye
from photographic plates.  It is only incomplete for very late Hubble types (T $>6$, i.e., later than Scd; \citealt{dejo}).  Values of $V_{\rm flat}$ 
were determined from optical and radio observations.  For the optical data, $V_{\rm flat}$ was
defined as half the difference  between the maximum and minimum velocities of the H$\alpha$ rotation curves.
For the radio data, $V_{\rm flat}$ was defined as half the width of the HI profile between points where the intensity
falls to 50\% of the highest values; these values were then corrected for dispersion and converted to optical rotation velocities
by multiplying by 1.03 and then subtracting 11 km s$^{-1}$ \citep[see \S3.4 and Figure 5 of][]{mat92}.
Disc half light radii, which are the result of $I$-band bulge-disc
decompositions from \citet{dejo}, are converted to disc scale lengths by dividing by 1.679 (the exact ratio of the half-mass
radius to the scale radius for a pure exponential disc).
Only those galaxies with rotation velocities greater than 125 km s$^{-1}$ are used.  This helps us to avoid galaxies
with rotation curves which do not flatten out at the radii measured.  The distribution of galaxies in $j$ 
versus $V_{\rm flat}$ does not differ significantly from the galaxy sample commonly used in the literature \citep{cour07}, 
but it has a better completeness.

To address the second need, and relate galaxies to the dark matter haloes in Figure~\ref{fig:haloes_only}, 
we assume for simplicity that the characteristic rotation velocity of a galaxy (which we take to be $V_{flat}$)
and that of its host halo at $R_{200}$ are equal.
For a massless disc in a \citet{nfw} halo, $V_c$ at the location of the galaxy can be about half its value at $R_{200}$.
However, the self-gravity of the baryons is expected to increase $V_c$ in the inner parts of haloes.  The amount by which it increases
is difficult to calculate theoretically, so we look to observations.  \citet{dutt} find a very small conversion factor
between $V_c$ at $R_{200}$ and at the location of galaxy discs.  In their analysis, \citet{dutt}
combined dark halo masses measured from satellite kinematics and
weak gravitational lensing to show that $V_{2.2} \simeq V_{200}$ for $V_{2.2} = 90 - 260$ km s$^{-1}$,
where $V_{2.2}$ is the galaxy rotation velocity measured at 2.2 $I$-band scale lengths.  
This equivalence is also consistent with semi-analytic models of galaxy formation which require a 
similar ratio between galaxy and halo velocities to simultaneously match the local Tully-Fisher relation and 
galaxy luminosity function \citep[e.g.,][and references therein]{dvdb}.

In Figure~\ref{fig:haloes_gals}, we compare the distribution of $j$ versus $V_{\rm flat}$ for galaxies described in this
section with the distributions of $j_{200}$, $j_{2000}$, and $j_{20,000}$ versus $V_{200}$ for 
dark matter haloes from Figure~\ref{fig:haloes_only}.  As discussed above, it is assumed that haloes have
the same rotation velocities as the galaxies they host, so they can be directly compared in 
Figure~\ref{fig:haloes_gals}.  The halo relations from Figure~\ref{fig:haloes_only} for $V_{20,000}$ are not shown because they
are not significantly different from those for $V_{200}$.
We fit a linear relation to the galaxies using 100 
bootstrap re-samplings and a generalised least squares fitting routine \citep{wei2}, which gives a slope 
of $2.5\pm0.1$ rms.  We also fit a linear relation to the haloes  in Figure~\ref{fig:haloes_gals} 
for $j_{200}$ versus $V_{200}$ for circular velocities which span the velocity range of the galaxies, $125 < V_{200} < 315$ km s$^{-1}$.  
This results in a slope of $1.92 \pm 0.02$ rms.  The distribution of galaxies 
has a similar slope to that of the haloes, as found by \citet{fall83} and others \citep[e.g.,][]{mmw, stein},
and approximately half the average rms scatter (0.15 dex versus 0.27 dex).   The lower scatter compared to the haloes
is related to the finding by \citet{dejo} that the width of the observed scale-radius distribution of galactic discs
is narrower than that  expected from the distributions of halo spin parameters in cosmological simulations.
For the halo $j_{2000}$ 
versus $V_{200}$ and $j_{20,000}$ versus $V_{200}$ relations, the slopes are $1.80 \pm 0.02$ and $1.26 \pm 0.02$, respectively,
and the average rms scatters are 0.28 and 0.26, respectively.
The slopes flatten slightly with increasing $\Delta$, but the scatter remains constant to within errors.
Given all the factors not included in our simple picture, we consider it remarkable how similar the galaxy
and halo slopes are.  The main result of this paper is encapsulated in the much larger difference in normalisation between
galaxies and haloes.  We choose to measure this difference at approximately the center of the distributions, at log $V_{rot}$=2.35 ($V_{rot}= 224$ km s$^{-1}$).  
The average  normalisation of the galaxies is less than that of the haloes
for $j_{200}$ by a factor of $\sim 2$ (0.30 dex), consistent with previous studies \citep[e.g.,][]{stein,dutt11}.
The average normalisation of the galaxies is greater than that of the haloes for
$j_{2000}$ and $j_{20,000}$ by factors of $\sim 2$ (0.30 dex) and $\sim 5.6$ (0.75 dex), respectively.  

We quantify the dependence of $j_{\Delta}$ on $\Delta$ as follows.
We start by measuring halo $j$ for a range of $\Delta$ and compare them with $j$ measured for
the galaxy sample, as in Figure~\ref{fig:haloes_gals}.  In Figure~\ref{fig:offset}, we show
the average difference between galaxy and halo $j$ (measured at log $V_{rot}= 2.35$) as a function of $\Delta$,
halo outer radius in terms of $R_{200}$ (i.e., $R_{\Delta}/R_{200}$), 
and halo mass in terms of $M_{200}$ (i.e., $M_{\Delta}/M_{200}$).  
The quantities $R_{\Delta}/R_{200}$ and $M_{\Delta}/M_{200}$ are
average values for all haloes with circular velocities which span $125 < V_{200} < 315$ km s$^{-1}$.
The value of $\Delta$ at which the average $j$ of galaxies and 
haloes match (i.e., $<{\rm log} \ j_{\rm galaxies} - {\rm log} \ j_{\Delta}> \ =  0$) 
is $578^{+34}_{-31}$.  The halo $j_{578}$ versus $V_{200}$ relation has a slope of $1.89 \pm 0.03$
and an average rms scatter of 0.28 dex over the velocity range of the galaxies.
This value of $\Delta$ corresponds to $R_{BC}=R_{\Delta=578}/R_{200} = 0.63^{+0.2}_{-0.1}$ and $ M_{\Delta=578}/M_{200} = 0.74\pm0.1$.  
We calculate these values by interpolating the curves in Figure~\ref{fig:offset},
and the errors by considering the limiting case that galaxy $j$ values
are systematically over and under-estimated by the assumed measurement uncertainty.
If baryons conserved $j$ perfectly during galaxy formation, then the collapse radius
$R_{BC}$ is $\simeq 63$\% of the virial radius $R_{200}$.
This portion of the haloes contains on average 74\% of their mass,
and if baryons and dark matter are initially well-mixed, the same percentage of the baryons.
However, as discussed in the next section, this radius and mass fraction are probably not simple step functions;
therefore we regard them as ``effective" quantities.

\section{Discussion}

In this paper, we determine the extent to which the approximate factor of 2 discrepancy between the $j$ of 
galaxies and their expected host dark matter haloes is sensitive to the conventional assumption that $R_{BC} = R_{200}$.
This difference in $j$ is usually attributed to loss of baryonic $j$ during galaxy formation.
However, there is no physical reason for the assumption that these radii are equal to at least within a factor of $\sim2$,
as emphasized by \citet{fall2}.  This is because different physics governs each, namely dissipational and dissipationless 
physics for $R_{BC}$  and $R_{Vir}$, respectively.  The only constraint on the relationship between
these radii is that $R_{BC}$ must be interior to $R_{Vir}$
since baryons cannot collapse from a region that is not incorporated into the halo. 
A $R_{BC}$ which is interior to $R_{Vir}$ is a natural expectation in the standard 
theory of galaxy formation where the inner parts of haloes collapse first.  
As $R_{BC}$ decreases, the discrepancy between the $j$ of galaxies and haloes is alleviated.
We show that the discrepancy can be explained entirely by 
a $R_{BC}$ which is $\sim60$\% of $R_{Vir}$.  

To do so, we determine the value of $R_{BC}$ at which the $j$ of galaxies and haloes match.  This is done by comparing the distribution of $j$ 
observed for a sample of local disc galaxies, for which the completeness is understood, to that predicted 
for their host dark matter haloes from a dark matter-only simulation of the Universe.  It is assumed that 
galaxies and haloes can be related directly via their rotation velocities.  The necessary value of the density contrast $\Delta$
needed to define the haloes which have the same average $j$ as galaxies is
$\sim 600$.  This corresponds to an average effective $R_{BC}$
which is $\sim 60$\% of $R_{200}$, and an average halo mass which is $\sim 75$\% of $M_{200}$.  Therefore, if galaxies formed from baryons
initially present in the inner parts of their host haloes and conserved $j$ perfectly, 
the baryons would come from within $R_{BC}$ and would comprise this percentage of the baryons in the halo.

Even under the assumption of perfect conservation of $j$, $R_{BC}$ is not likely a sharp boundary. 
The baryons which form the galaxy may only on average come from within
$R_{BC}$, with most material originating from smaller radii, but some from more distant radii.
In addition, the smaller scatter of the galaxies in $j$ versus $V$ compared to that of the haloes
may indicate a mechanism by which only  selected baryons form the disc, regulatory processes which act upon the baryons,
and/or haloes which form non-disc galaxies.  
This is because, in our simple picture, the initial distribution of baryons in $j$ versus $V$ is expected to mirror that of the dark matter.
Therefore, if only selected baryons formed discs or regulatory processes acted upon them during disc formation,
it may be expected that the baryons which form the discs would have a narrower distribution in $j$ versus $V$.
In addition, since we compare the predicted properties of dark matter haloes with those of
disc galaxies, not ellipicals which rotate slower than discs, it stands to reason that the combined population
 of discs and ellipticals would be broader in $j$ versus $V$ \citep{fall83}.

Eventually, it should be possible to compute $R_{BC}$ from hydrodynamical and dark matter simulations of galaxy formation 
in a cosmological context.  Current simulations may have spatial and mass resolutions that are
too coarse to model accurately the complex processes expected to be at play, such as gas shocks, cloud-cloud collisions, and a
multiphase medium.  These processes affect the rate at which the baryons collapse, but they have may relatively little influence on the  
angular momentum of the resulting galactic discs.

A number of phenomena can alter the $j$ of galaxies (see \citealt{fall2} and \citealt{roman} for more complete discussions
of these phenomena).  For example, torques exerted between the dark matter and the baryons
could in principle spin up the halo and spin down the disc.  Minor mergers might also affect the $j$ of galaxies.
In addition, feedback from star-formation can alter $j$ differently depending on how
it varies with radius.  Material in outflows may be launched from inner
or outer radii, or both.  If material is primarily removed from the inner or outer parts of galaxies, galaxy
$j$ will increase or decrease, respectively.  If feedback is active but independent of radius, then
there would be no change in $j$.  We expect some of these phenomena to alter the $j$ of discs, but whether
they have a major or a minor effect on galaxy $j$ is still uncertain.
In order to perform a more detailed comparison of 
galaxies and haloes, we need a better understanding of the processes of $j$ transfer
in galaxy formation, and whether outflows can change the $j$ of galaxies.


In summary, the difference between the predicted initial and the observed final $j$ of 
galaxies, which is conventionally attributed solely to angular momentum loss, hinges on the loosely motivated
assumption that all the baryons within $R_{Vir}$ collapse to form galaxies.  There is no
physical reason why this has to be the case.  If baryons in the inner parts of haloes collapse first,
as is expected, then the $j$ discrepancy between galaxies and haloes can be fully explained by a collapse
radius $R_{BC}$ which is $\sim60$\% of the virial radius $R_{Vir}$.  In the future, baryons from progressively
larger radii in the halo may collapse, and at some point in time $R_{BC}$ might equal $R_{Vir}$.
In reality, it may be that a combination of a preference of collapse of the inner parts and some $j$ transfer 
between baryons and dark matter is needed to solve the problem.



\section*{Acknowledgments}
This research was supported by an appointment to the NASA Postdoctoral Program at
NASA's Goddard Space Flight Center, administered by Oak
Ridge Associated Universities through a contract with NASA.
The research of JD is partly supported by Adrian Beecroft, the Oxford Martin School and STFC.
This work was performed using code and simulations from the Horizon collaboration (http://www.projet-horizon.fr).
S.A.K. and J. D. are grateful to David S. Graff.   We thank Aaron Dutton and Aaron Romanowski for helpful
comments on drafts of this paper.

\end{document}